\documentclass[12pt]{iopart}
\usepackage{amssymb}
\usepackage{euscript}
\usepackage[english]{babel}

\usepackage{color}

\newcommand{\eqref}[1]{(\ref{#1})}

\begin{document}

\title[Lax representations and integrable hierarchies via exotic cohomology. II]{%
Lax representations with non-removable parameters and integrable hierarchies  of PDEs via exotic cohomology
of symmetry algebras. II}

\author[O.I. Morozov]{Oleg I. Morozov}

\vskip 10 pt

\address{
  Faculty of Applied
  Mathematics, AGH University of Science and Technology,
  \\
  Al. Mickiewicza 30,
  Cracow 30-059, Poland,
    \\
   e-mail: morozov\symbol{64}agh.edu.pl
}



\begin{abstract}
We derive new Lax representations for the hyper-CR equation of Ein\-stein--Weyl structures and
for the associated integrable hierarchy.
\end{abstract}

\maketitle

\section{Introduction}

In this paper we continue to study the relations between structure of symmetry algebras and Lax
representations of integrable partial differential equations \cite{Morozov2017,Morozov2018,Morozov2019}.
We consider the hyper-CR equation for Ein\-stein--Weyl structures
 \cite{Kuzmina1967,Mikhalev1992,Pavlov2003,Dunajski2004}
\begin{equation}
u_{yy} = u_{tx} + u_y\,u_{xx} - u_x\,u_{xy},
\label{Pavlov_eq}
\end{equation}
and generalize the Lax representation
\begin{equation}
\left\{
\begin{array}{lcl}
v_t &=& (\lambda^2-\lambda\,u_x-u_y)\,v_x,
\\
v_y &=& (\lambda-u_x)\,v_x,
\end{array}
\right.
\label{Pavlov_eq_covering}
\end{equation}
thereof.
In \cite{Morozov2019}  we show that
(\ref{Pavlov_eq_covering}) can be inferred from nontriviality of the second exotic cohomology group of
$\mathrm{Sym}_0(\EuScript{E})$. A nontrivial cocycle provides an extension of the symmetry algebra
$\mathrm{Sym}_0(\EuScript{E})$. The linear combination of the Maurer--Cartan ({\sc mc}) form $\sigma$
of the extension and certain horizontal {\sc mc} form of $\mathrm{Sym}_0(\EuScript{E})$ gives the
Wahlquist--Estabrook form that generates the Lax representation (\ref{Pavlov_eq_covering}).
In the section \ref{new_Lax_representation} of the
present paper we find new Lax representation for equation (\ref{Pavlov_eq}). We consider the linear
combination of $\sigma$ with arbitrary basic horizontal {\sc mc} form of $\mathrm{Sym}_0(\EuScript{E})$
and find conditions under which this combination produces a Lax representation for equation
(\ref{Pavlov_eq}). As a result we generalize the the Lax representation (\ref{Pavlov_eq_covering}).

Expansion of (\ref{Pavlov_eq_covering}) into the Taylor series w.r.t. $\lambda$ yields an
infinite-dimensional Lax representation for (\ref{Pavlov_eq}).  This infinite-dimensional Lax
representation can be used to study nonlocal symmetries \cite{BaranKrasilshchikMorozovVojcak2018}
and nonlocal conservation laws \cite{MakridinPavlov2017} of equation (\ref{Pavlov_eq}). In section
\ref{expansion_section} we expand the new Lax representation into the Taylor series and find explicit
expressions for the coefficients of the obtained infinite-dimensional Lax representation. This Lax
representation can be considered as a pair of hydrodynamic chains (see \cite{Pavlov2003} and
references therein) whose compatibility conditions coincide with (\ref{Pavlov_eq}).

In section \ref{hierarchy_section} we derive  new  Lax representation
(\ref{Pavlov_eq_hierarchy_new_covering}) for the integrable hierarchy associated with equation
(\ref{Pavlov_eq}).

We follow definitions and notation of
\cite{KrasilshchikVerbovetsky2011,KrasilshchikVerbovetskyVitolo2012,Morozov2019},
see also \cite{KrasilshchikVinogradov1984,KrasilshchikVinogradov1989,VK1999}.

\section{The generalized Lax representation}
\label{new_Lax_representation}
As it was shown in \cite{Morozov2019}, the structure equations for the symmetry algebra
$\mathrm{Sym}_0(\EuScript{E})$ of (\ref{Pavlov_eq}) have the form
\begin{equation}
\left\{
\begin{array}{lcl}
d\alpha_0 &=& 0,
\\
d\alpha_1 &=& \alpha_0 \wedge \alpha_1,
\\
d \Theta &=& \nabla_1 (\Theta) \wedge \Theta + (h_0\,\alpha_0 + h_0^2\,\alpha_1) \wedge \nabla_0 (\Theta),
\end{array}
\right.
\label{structure_equations_for_Pavlov_eq}
\end{equation}
where
\begin{equation}
\Theta = \sum \limits_{k=0}^{3} \sum \limits_{m=0}^{\infty} \frac{h_0^kh_1^m}{m!}\,\theta_{k,m},
\label{Theta_Pavlov}
\end{equation}
where $h_0$ and $h_1$ are formal parameters such that $dh_i=0$ and $h_0^k =0$ when $k>3$, $\nabla_0$ is
the derivative with respect to $h_0$ in $\mathbb{R}_3[h_0] = \mathbb{R}[h_0]/(h_0^4=0)$,
$\nabla_1 =\partial_{h_1}$, while $\alpha_0$, $\alpha_1$, and $\theta_{k,m}$ are {\sc mc} forms of
$\mathrm{Sym}_0(\EuScript{E})$. The second exotic cohomology group
$H^2_{c\,\alpha_0}(\mathrm{Sym}(\EuScript{E}))$ is nontrivial when $c=1$, the nontrivial 2-cocycle
$\alpha_0 \wedge \alpha_1$ of the differential $d_{\alpha_0}$ defines a non-central
extension  of the Lie algbera $\mathrm{Sym}(\EuScript{E})$. The additional Maurer--Cartan form
$\sigma$ for the extended Lie algebra is a solution to
$d_{\alpha_0} \sigma = \alpha_0 \wedge \alpha_1$, that is, to equation
\begin{equation}
d \sigma = \alpha_0 \wedge \sigma + \alpha_0 \wedge \alpha_1.
\label{extension_se_1}
\end{equation}
This equation is  compatible with the structure equations \eqref{structure_equations_for_Pavlov_eq}
of the Lie algebra $\mathrm{Sym}_0(\EuScript{E}_1)$.

We find the following expressions for the {\sc mc} forms:
$\alpha_0 = dq$,
$\alpha_1 = - \e^q\, ds$,
$\sigma  = \e^q\,(dv-q \,ds)$,
$\theta_{0,0} = r\,dt$,
$\theta_{1,0} = r\, \mathrm{e}^q\,(dy - (u_x-2\,s)\,dt)$,
$\theta_{2,0}= r\, \mathrm{e}^{2q}\,(dx -(u_x-s)\,dy - (u_y+s\,u_x-s^2)\,dt)$,
$\theta_{3,0} = r\, \mathrm{e}^{3q}\,(du - u_t\,dt - u_x\,dx -u_y\,dy)$,
where $q$, $s$, $v$, and $r$  are free parameters.
As it was shown in \cite{Morozov2019}, the form $\sigma - \theta_{2,0}$ is the Wahlquist--Estabrook form
of the Lax representation (\ref{Pavlov_eq_covering}).

Consider the linear combination
\[
\sigma - \sum \limits_{k=0}^2 c_k\,\theta_{k,0} =
\mathrm{e}^q\,\left(
dv - q\,ds - c_2\,r\,\mathrm{e}^q\,dx
-r\,(c_1+c_2\,\mathrm{e}^q (s-u_x))\,dy
\right.
\]
\[
\qquad\left.
-r\,(c_0\,\mathrm{e}^{-q}+c_1\,(2\,s-u_x)+c_2\,\mathrm{e}^q (s^2-s\,u_x-u_y))\,dt
\right)
\]
of the form $\sigma$ and the {\it basic horizontal forms} $\theta_{0,0}$, $\theta_{1,0}$, $\theta_{2,0}$.
Put without loss of generality $c_2=1$ and rename $q=v_s$, $r= v_x\,\exp\, (-v_s)$.
Then the form
\[
\sigma - c_0\,\theta_{0,0} - c_1\,\theta_{1,0} - \theta_{2,0}=
\mathrm{e}^q\,\left(
dv - v_s\,ds - v_x\,(dx+(s-u_x+c_1\,\mathrm{e}^{-v_s})\,dy
\right.
\]
\[
\qquad\left.
+(s^2-s\,u_x-u_y+c_1\,\mathrm{e}^{-v_s}\,(2\,s-u_x)+c_0\,\mathrm{e}^{-2v_s})\,dt)
\right)
\]
is equal to zero whenever there hold
\begin{equation}
\left\{
\begin{array}{lcl}
v_t &=& \left(s^2-s\,u_x-u_y+c_1\,\mathrm{e}^{-v_s}\,(2\,s-u_x)+c_0\,\mathrm{e}^{-2v_s}\right)\,v_x,
\\
v_y &=& \left(s-u_x+c_1\,\mathrm{e}^{-v_s}\right)\,v_x.
\end{array}
\right.
\label{generalized_covering}
\end{equation}
The compatibility condition $(v_t)_y=(v_y)_t$ reads
\begin{equation}
(c_1^2-c_0)\,v_x\,\mathrm{e}^{-2 v_s} \,(2\,v_x+u_{xx})
- v_x\,(u_{yy}-u_{tx} - u_y\,u_{xx}+u_x\,u_{xy}) =0.
\label{compatibility_of_generalized_covering}
\end{equation}
In Appendix we prove that system (\ref{generalized_covering}) does not define a Lax representation
for equation (\ref{Pavlov_eq}) when $c_0 \neq c_1^2$. Consider case $c_0 = c_1^2$. When $c_1=0$, we get
the Lax representation (\ref{Pavlov_eq_covering})  with $\lambda=s$. When $c_1 \neq 0$ we put $c_1=1$
without loss of generality. This yields the new Lax representation
\begin{equation}
\left\{
\begin{array}{lcl}
v_t &=& \left(s^2-s\,u_x-u_y+\mathrm{e}^{-v_s}\,(2\,s-u_x)+\mathrm{e}^{-2v_s}\right)\,v_x,
\\
v_y &=& \left(s-u_x+\mathrm{e}^{-v_s}\right)\,v_x
\end{array}
\right.
\label{new_covering}
\end{equation}
of equation (\ref{Pavlov_eq}).


\section{Infinite-dimensional Lax representation}
\label{expansion_section}

For the infinite sequence $\underline{a} = (a_1, a_2, ..., a_n, ...)$ put $R_0(\underline{a}) =1$ and
\[
R_k(\underline{a}) = S_k\left(-\frac{a_1}{1}, \dots, -\frac{a_{m}}{m!}, \dots, -\frac{a_{k}}{k!}\right)
\]
for $ k \ge 1$, where the {\it elementary Schur polynomials}  $S_k$ are defined by the generating
function
\[
\exp\,\left(\sum \limits_{m=1}^{\infty} a_m \,z^m\right) = 1 + \sum \limits_{k=1}^\infty S_k(a_1, \dots , a_k) \,z^k.
\]
For expansion
\[
v= \sum \limits_{k=0}^{\infty} \frac{v_k\,s^k}{k!},
\qquad v_k=v_k(t,x,y),
\]
of function $v$ into the Taylor series w.r.t. $s$ denote $\underline{v} = (v_2,v_3, \dots, v_m, \dots)$.
Then we have
\[
v_s= \sum \limits_{k=0}^{\infty} \frac{v_{k+1}\,s^k}{k!},
\]
and
\begin{equation}
\mathrm{e}^{-v_s} = \mathrm{e}^{-v_1}\,\left(\sum \limits_{k=0}^\infty
R_k(\underline{v}) \,s^k\right).
\label{evs_expansion}
\end{equation}
Substituting for (\ref{evs_expansion}) into (\ref{new_covering}) yields the infinite-dimensional
Lax representation
\[
\fl
\left\{
\begin{array}{lcl}
v_{0,t} &=&
\displaystyle{
-\left(u_y +u_x\,\mathrm{e}^{-v_1}+\mathrm{e}^{-2\,v_1}\right)\,v_{0,x},
}
\\
v_{1,t} &=&
\displaystyle{
\left(v_2\,\mathrm{e}^{-v_1}\,\left(u_x-2\,\mathrm{e}^{-v_1}\right)+2\,\mathrm{e}^{-v_1}-u_x\right)\,v_{0,x}
+\left(\mathrm{e}^{-2 v_1}-\mathrm{e}^{-v_1}\,u_x-u_y\,\right)\,v_{1,x},
}
\\
v_{k,t} &=&
\displaystyle{
k\,(k-1)\,v_{k-2,x}- k \,u_x\,v_{k-1,x}-u_y\,v_{k,x}
+2\,\mathrm{e}^{-v_1}\,\sum \limits_{m=0}^{k-1}
R_{k-m-1}(\underline{v})
\,\frac{k!}{m!}\,v_{m,x}
}
\\
 &&
\displaystyle{
+\mathrm{e}^{-v_1}\,\sum \limits_{m=0}^k
\left(\mathrm{e}^{-v_1}\,R_{k-m}(2\,\underline{v})-u_x\,R_{k-m}(\underline{v})\right)
\,\frac{k!}{m!}\,v_{m,x},
}
\\
v_{0,y} &=&
\displaystyle{
\left(-u_x+\mathrm{e}^{-v_1}\right)\,v_{0,x},
}
\\
v_{1,y} &=&
\displaystyle{
\left(1-v_2\,\mathrm{e}^{-v_1}\right)\,v_{0,x}
+
\left(-u_x+\mathrm{e}^{-v_1}\right)\,v_{1,x},
}
\\
v_{k,y} &=&
\displaystyle{
k\,v_{k-1,x}-u_x\,v_{k,x}
+
\mathrm{e}^{-v_1}\,
\sum \limits_{m=0}^{k}
R_{k-m}(\underline{v})
\,\frac{k!}{m!}\,v_{m,x}.
}
\end{array}
\right.
\]
with $k \ge 2$.
This Lax representation is a generalization of the `positive' covering of (\ref{Pavlov_eq})
from \cite[\S~4.1]{BaranKrasilshchikMorozovVojcak2018} and
can be considered as a pair of hydrodynamic chains whose compatibility
conditions coincide with (\ref{Pavlov_eq}).


\section{Lax representation for the associated hierarchy}
\label{hierarchy_section}

The symmetry algebra $\mathrm{Sym}_0(\EuScript{E})$ of equation (\ref{Pavlov_eq}) admits an increasing
sequence of natural extensions
$\mathrm{Sym}_0(\EuScript{E}) = \mathfrak{p}_4 \subsetneq \mathfrak{p}_5 \subsetneq \dots \subsetneq
\mathfrak{p}_{n} \subsetneq \mathfrak{p}_{n+1} \subsetneq \dots$, where the Lie algebra
$\mathfrak{p}_{n+1}$ has the same structure equations (\ref{structure_equations_for_Pavlov_eq})  with
(\ref{Theta_Pavlov}) replaced by
\begin{equation}
\Theta = \sum \limits_{k=0}^{n} \sum \limits_{m=0}^{\infty} \frac{h_0^kh_1^m}{m!}\,\theta_{k,m},
\label{Theta_Pavlov_n}
\end{equation}
and $h_0^k = 0$ for $k \ge n+1$.
The nontrivial 2-cocycle $\alpha_0 \wedge \alpha_1$ of  $d_{\alpha_0}$ defines a non-central extension
$\widehat{\mathfrak{p}}_{n+1}$ of the Lie algebra $\mathfrak{p}_{n+1}$. The structure equations for
$\widehat{\mathfrak{p}}_{n+1}$ are given by \eqref{structure_equations_for_Pavlov_eq},
\eqref{Theta_Pavlov_n}, and \eqref{extension_se_1}.

As we show in \cite{Morozov2019}, the basic horizontal forms $\theta_{0,0}$, ... , $\theta_{n-1,0}$
for the Lie algebra $\widehat{\mathfrak{p}}_{n+1}$ with fixed $n>4$ can be expressed as follows.
Put $p_0=1$ and for $i \ge 0$, $j \in \{0, \dots, i\}$ define polynomials
$P_{ij}=P_{ij}(s)$
of variable $s$ by the formula
\begin{equation}
P_{ij} = \sum \limits_{k=0}^{j} (-1)^k\,
\left(
\begin{array}{c}
i-j+k-1\\k
\end{array}
\right)
\,p_{j-k}\,s^k.
\label{P_definition}
\end{equation}
Coefficients of $P_{ij}$ depend on parameters $p_1$, ..., $p_j$.
We have
\begin{equation}
\theta_{k,0} = \e^{kq}\,r\,\sum \limits_{j=0}^{k} P_{kj}\,dx_{n-k+j-1}
\label{theta_k_0}
\end{equation}
for  $k \in \{0, \dots, n\}$.
Then we put $x_{-1} =u$ and  enforce $\theta_{n,0}$ to be the contact form:
\[
\theta_{n} = \e^{n q} r\,\left(du - \sum \limits_{i=0}^{n-1} u_{x_i} dx_i\right).
\]
This requirement yields  the linear triangular system
\begin{equation}
P_{k,k-i} = - u_{x_{k-i-1}}, \qquad i \in \{0, \dots, k-1\},
\qquad k \in \{1, \dots, n\}
\label{p_system}
\end{equation}
with unknows $p_1$, $p_2$, ... , $p_n$. The final expressions for forms $\theta_{k.0}$ are obtained
by substituting for the solution of (\ref{p_system}) into (\ref{theta_k_0}).

As it was shown in \cite{Morozov2019},  form  $\sigma - \theta_{n-1}$ is the Wahlquist--Estabrook form for the
Lax representation of the $n$-th element of the inegrable hierarchy associated to equation
(\ref{Pavlov_eq}). To generalize this result we consider the linear combination
\[
\sigma - \sum \limits_{k=0}^{n-1} c_k\,\theta_{k,0}.
\]
where $c_{n-1} = 1$ without loss of generality. This form is equal to zero whenever an overdetermined
system of {\sc pde}s  for function $v=v(x_0, \dots , x_{n-1}, s)$ holds. The compatibility conditions
of this system coincide with equation (\ref{Pavlov_eq}) when either $c_i =0 $ for
$i \in \{0, \dots, n-2\}$, that is, we have the Wahlquist--Estabrook form $\sigma- \theta_{n-1}$ from
\cite{Morozov2019},  or $c_{k} = c_{n-2}^{n-k-1}$ for $k\in \{0,\dots, n-3\}$. In the last case we put
$c_{n-2}=1$ without loss of  generality and obtain
\begin{equation}
\fl
\left\{
\begin{array}{lcl}
v_{x_1} &=& (s-u_{x_0})\,v_{x_0},
\\
v_{x_2} &=& v_{x_0} \,\mathbb{D}_2(s^2-s\,u_{x_0}-u_{x_1}),
\\
&& \dots
\\
v_{x_i} &=&v_{x_0} \,\mathbb{D}_i\left(s^i -\sum \limits_{j=0}^{i-1}\,s^{i-j-1}\,u_{x_j}\right),
\\
&& \dots
\\
v_{x_{n-1}} &=&v_{x_0} \,\mathbb{D}_{n-1}\left(s^{n-1} -s^{n-2} \,u_{x_0} -s^{n-3} \,u_{x_1}
- \dots- s\,u_{x_{n-3}} - u_{x_{n-2}}
\right)
\end{array}
\right.
\label{Pavlov_eq_hierarchy_new_covering}
\end{equation}
with differential operators
\[
\mathbb{D}_m = \sum \limits_{k=0}^m \frac{\mathrm{e}^{-k\,v_s}}{k!}\,\frac{\partial^k}{\partial s^k}.
\]
The compatibility conditions of this system give the $n$-th element of the integrable hierarchy
\begin{equation}
u_{x_{m}x_{k}}=u_{x_{m-1}x_{k+1}}+u_{x_k}u_{x_0x_{m-1}}-u_{x_{m-1}}u_{x_0x_{k}},
\label{hierarchy_equations}
\end{equation}
$m \in \{1, \dots, k\}$, $k \in \{1, \dots, n-2\}$, associated with equation (\ref{Pavlov_eq}).
When $n=3$, the change of notation $x_0 \mapsto x$, $x_1 \mapsto y$, $x_2 \mapsto t$ in
(\ref{Pavlov_eq_hierarchy_new_covering}) and (\ref{hierarchy_equations}) gives (\ref{new_covering}) and
(\ref{Pavlov_eq}).

\section*{Acknowledgments}

This work was partially supported by the Faculty of Applied Mathematics of AGH UST statutory tasks within
subsidy of Ministry of Science and Higher Education.

It it my great pleasure to thank the Arctic University of Norway --- University of Troms{\o} for  the
financial support of my visit to Troms{\o}, where a part of the work on was performed. My special thanks
are to professor Boris Kruglikov for his kind help in organizing my visit and for very fruitful
discussions.

\section*{Appendix}

When $c_1^2\neq c_0$, equation (\ref{compatibility_of_generalized_covering}) yields either $v_x=0$, as
so $v_t =v_y =0$,  or $v_x = -\frac{1}{2}\,u_{xx}$. In the last case we have $v_{xs}=)$, and therefore
\[
0 = v_{xys}  = - v_{xx}\, \mathrm{e}^{-v_s}\,\left(c_1\,v_{ss} - \mathrm{e}^{-v_s}\right).
\]
If there holds $c_1\,v_{ss} - \mathrm{e}^{-v_s}\neq 0$, then $v_{xx}=0$, and
\[
0 = v_{txs} = -\frac{1}{2}\,v_x^2\,\mathrm{e}^{-v_s}\,\left(c_1\,v_{ss} - \mathrm{e}^{-v_s}\right)
\]
gives $v_x = 0$ again, while $c_1\,v_{ss} - \mathrm{e}^{-v_s} = 0$ entails
\[
0 = v_{txs} = \frac{2}{c_1}\,(c_1^2-c_0)\,v_{xx}\,\mathrm{e}^{-v_s}
\]
and so $v_{xx}= -\frac{1}{2}\,u_{xxx} =0$, or
\begin{equation}
u=W_2(t,y)\,x^2 + W_1(t,y)\,x+W_0(t,y).
\label{u_substitution_1}
\end{equation}
Then from the second equation of (\ref{new_covering}) we have
$v_{xy} = -v_x\,u_{xx}$, or
\begin{equation}
u_{xxy} = -u_{xx}^2.
\label{u_xxy_eq}
\end{equation}
Substituting for (\ref{u_substitution_1}) into (\ref{Pavlov_eq}) and (\ref{u_xxy_eq}) gives a family
\begin{equation}
u=(A_2\,y+A_1)\,x+\frac{1}{6}\,(A_2^{\prime}-A_2^2)\,y^3+\frac{1}{2}\,(A_1^{\prime} -A_1\,A_2)\,y^2
+A_3\,y+A_4,
\label{u_substitution_2}
\end{equation}
of solutions to equation (\ref{Pavlov_eq}) that include arbitrary functions  $A_1=A_1(t)$, ... ,
$A_4=A_4(t)$. For a solution from  (\ref{u_substitution_2}) we have $u_{xx}$, so $v_x = v_t=v_y=0$ again.

Note that (\ref{u_substitution_2}) is a subset of the family of solutions to (\ref{Pavlov_eq}) that are
defined by restriction $u_{xxxx} = 0$ and have the form
\[
u=A_0(t)\,x^3+W_2(t,y)\,x^2 + W_1(t,y)\,x+W_0(t,y),
\]
where there holds
\[
\left\{
\begin{array}{lcl}
W_{2,yy} &=& -2\,W_2\,W_{2,y}+3\,(A_0\,W_{1,y}+A_0^{\prime}),
\\
W_{1,yy} &=& 2\,(W_{2,t}-W_1\,W_{2,y}+3\,A_0\,W_{0,y}),
\\
W_{0,yy} &=& 2\,W_2\,W_{0,y}-W_1\,W_{1,y}+W_{1,t}.
\end{array}
\right.
\]

\end{document}